  \documentstyle[twocolumn,epsbox]{jpsj}

\def\mbf(#1){\mbox{\boldmath $#1$}}
\def\runtitle{Effect of Spin-orbit Impurity Scattering
}
\def\runauthor{Toshifumi {\sc Yagi} and Kazuhiro {\sc Kuboki} }

\title
{Effect of spin-orbit impurity scattering in the superconducting state 
of $t-J$ model}

\author{Toshifumi {\sc Yagi} and Kazuhiro {\sc Kuboki}\footnote{E-mail:
kuboki@phys.sci.kobe-u.ac.jp}}

\inst{Department of Physics, Kobe University, Kobe 657-8501, Japan}

\recdate{ }

\abst{  
We study the effect of magnetic impurities in the $d_{x^2-y^2}$-wave 
superconducting (SC) state of the two dimensional $t-J$ model.
The spin-orbit and the spin-exchange interactions are examined by treating 
the impurity as a classical spin.   
The Bogoliubov de Gennes equation derived within a slave-boson 
mean-field approximation is solved numerically at $T = 0$. 
The spin-exchange scattering induces spin-triplet $p$-wave SC order 
parameters near the impurity, while a SC state with broken time-reversal 
symmetry and a spontaneous current 
appears in the presence of the spin-orbit interaction. 
When both interactions coexist, it turns out that 
a state which carries a spontaneous spin current occurs. 
}

\kword{$d$-wave superconductivity, $t-J$ model, 
Bogoliubov de Gennes equation, magnetic impurity, 
spin current
}

\begin{document}
\sloppy
\maketitle

%\vfill\eject

The symmetry of the superconducting (SC) states in high-$T_{\rm c}$ 
cuprates has been studied intensively, and now it is established that 
the state has a predominantly $d_{x^2-y^2}$-wave character with a possible 
mixture of an $s$-wave component due to the orthorhombic lattice 
distortion in some systems.\cite{Scl}
Recently the effect of nonmagnetic (Zn) and Magnetic (Ni) impurities in 
high-$T_c$ superconductors are of particular interest both 
theoretically and experimentally. 
This is because the effects of randomness in such unconventional 
superconductors can be quite different from those in conventional 
superconductors.

The low-energy electronic states of high-$T_c$ cuprates can be 
described by the $t-J$ model on a square lattice\cite{PWA,ZR}.
Mean-field (MF) theories based on a slave-boson method predict a 
superconducting state with a $d_{x^2-y^2}$-symmetry.\cite{SHF,KL} 
They may also explain the magnetic\cite{TKF,FKNT} as well as 
the transport\cite{NL} properties of these systems 
if the gauge fields representing the fluctuations around the MF 
solutions are taken into account. 
The effect of nonmagnetic impurities in the SC state of this model 
has been examined by several authors. \cite{Tanaka,Tsuchi1,Ting1}

In this article we study the effect of a magnetic impurity in the 
superconducting state of the $t-J$ model by taking into account both
the spin-exchange and the spin-orbit interactions between 
electrons and an impurity spin. 
We consider the case of single impurity (located at a site 0),  
assuming that the concentration of impurities is low 
so that their effect can be treated independently. 
In this work the impurity spin is treated as classical, i.e., 
the Kondo effect will not be considered.  

The Hamiltonian of our system is 
\begin{equation}\begin{array}{rl}
H & =  \displaystyle -\sum_{\langle i,j \rangle,\sigma}
t_{ij}\big({\tilde c}_{i\sigma}^\dagger {\tilde c}_{j\sigma} + h.c. \big) \\
& \\
& +  \displaystyle \sum_{\langle i,j \rangle} 
J_{ij} {\vec S}_i\cdot{\vec S}_j 
- \mu\sum_{i\sigma}{\tilde c}_{i\sigma}^\dagger {\tilde c}_{i\sigma} \\
& +  \displaystyle V_0 \sum_\sigma 
{\tilde c}_{0\sigma}^\dagger{\tilde c}_{0\sigma} 
+ J_0 \sum_{\tau}{\vec S}_{imp}\cdot{\vec S}_{0+\tau} + H_{so}
\end{array}\end{equation}
where ${\tilde c}_{i\sigma}$ is an electron operator within the Hilbert space 
excluding double occupancy, and ${\vec  S}_i$ and $\mu$ are 
the spin-1/2 operator at a site $i$ and the chemical potential, respectively. 
For the transfer integrals $t_{ij}$ and 
the antiferromagnetic superexchange interaction $J_{ij}$, 
next-nearest-neighbor as well as nearest-neighbor terms are 
taken into account: 
%\begin{equation} 
%\begin{array}{rl}
$t_{ij} =   t \sum_\tau\delta_{i,j+\tau} 
+ t' \sum_\nu\delta_{i,j+\nu},  \ \
J_{ij} =   J \sum_\tau\delta_{i,j+\tau} 
+ J' \sum_\nu\delta_{i,j+\nu} $
%\end{array}
%\end{equation}
with $\tau = \pm {\hat x}, \pm {\hat y}$ and 
$\nu = \pm {\hat x} \pm {\hat y}$. 

The impurity part consists of three terms: 
(i) potential scattering $V_0$, which is assumed to be in the 
unitary limit ($V_0 \gg J, t$),  (ii) exchange interaction 
$J_0$ ($ \geq 0$) which acts between the impurity 
(${\vec S}_{imp}$; $S_{imp} = 1$) and its nearest-neighbor sites, 
and (iii)  spin-orbit interaction described by $H_{so}$. 
The model described by (i) and (ii) is essentially the same as the one 
studied by Poilblanc et al.\cite{Poil}. 
We simplify the model by treating ${\vec S}_{imp}$ as a classical spin, 
$\langle S^z_{imp} \rangle$.
The spin-orbit interaction $H_{so}$ is given by 
\begin{equation}
H_{so} = \displaystyle g \sum_\sigma \int d^2r \frac{1}{r^3}
S^z_{imp} {\tilde c}_\sigma^\dagger(r) [{\vec r}\times i{\vec \partial}]_z 
{\tilde c}_\sigma(r).  
\end{equation} 
in the continuum representation,  
and we transform this to that on a lattice by replacing the 
derivatives with discrete differences. 
In the actual calculations we take into account interactions only
up to second-neighbor sites of the impurity for simplicity of 
numerical calculations. 

We use the slave-boson method to enforce the condition of no 
double occupancy by introducing spinons ($f_{i\sigma}$; fermion) 
and holons ($b_i$; boson) (${\tilde c}_{i\sigma} = 
b_i^\dagger f_{i\sigma}$), and 
decouple this Hamiltonian by a mean-field approximation (MFA).
In the following we consider only the case of zero temperature ($T=0$), 
so that holons are Bose condensed.
Then the mean-field Hamiltonian is written in terms of spinons 
only, \cite{Gutz}
\begin{equation}
{\cal H}_{MFA} = \sum_i\sum_j \Psi_i^\dagger 
\left [\begin{array}{clcr}
W_{ij}^{(\uparrow)} & F_{ij} \\
F_{ji}^* & -W_{ji}^{(\downarrow)}
\end{array}\right ] 
\Psi_j
\end{equation} 
with  
\begin{equation}
\begin{array}{rl}
W_{ij}^{(\sigma)} = & \displaystyle 
-\big(t_{ij}\delta + \frac{1}{2}J_{ij}\chi_{ji}^{(-\sigma)}
+ \frac{1}{4}J_{ij}\chi_{ji}^{(\sigma)}\big)  \\ 
& \\ 
+ & \displaystyle 
\big[\frac{\sigma}{2}\big(\sum_\lambda 
J_{i,i+\lambda}\langle S^z_{i+\lambda} \rangle 
+ J_0 \langle S^z_{imp} \rangle \sum_\tau \delta_{i,0+\tau}\big) \\
- & \displaystyle \mu +V_0 \delta_{i,0}\big]\delta_{i,j},   \\
& \\
F_{ij} =& \displaystyle -\frac{1}{2}J_{ij}\Delta_{ij} 
- \frac{1}{4}J_{ij}\Delta_{ji},  \\
& \\ 
\Psi_i^\dagger = & \displaystyle
\big(f^\dagger_{i\uparrow}, f_{i\downarrow}\big)
\end{array}
\end{equation}
where $\delta$ is the doping rate, and the summations on $i$, $j$ 
and $\lambda$ are taken over all sites.
Here the SC order parameter (OP), $\Delta_{ij}$, the hopping OP, $\chi_{ij}$, 
and the magnetization, $\langle S^z_i \rangle$,  are defined as 
\begin{equation}
\begin{array}{rl}
& \Delta_{ij} = 
\langle f_{i\downarrow}f_{j\uparrow} \rangle, 
\ \ 
\chi_{ij}^{(\sigma)} = \langle f_{i\sigma}^\dagger f_{j\sigma}\rangle ,  \\
& \\
& \langle S^z_i \rangle = \langle f^\dagger_{i\uparrow}f_{i\uparrow} 
- f^\dagger_{i\downarrow}f_{i\downarrow} \rangle/2. 
\end{array}
\end{equation}  
The SCOP with the definite symmetry can be constructed from 
$\Delta_{ij}$. 
For example $d_{x^2-y^2}$-wave OP ($\Delta_d$) is defined as 
\begin{equation}
\Delta_d(i) = ({\tilde \Delta}_{i,i+x} + {\tilde \Delta} _{i,i-x} 
- {\tilde \Delta}_{i,i+y} - {\tilde \Delta}_{i,i-y})/4 
\end{equation}
with ${\tilde \Delta}_{ij} = (\Delta_{ij} + \Delta_{ji})/2$ being 
the singlet-pairing OP. 
Here we do not exclude the spin-triplet pairing, 
and they actually can be finite as will be seen later. 

In a uniform system without impurities only the $d_{x^2-y^2}$-wave OP 
is finite for the doping rate of our interest, i.e., $0.15 < \delta <  0.2$ 
(optimum and overdoped cases). 
In the presence of an impurity, however, OP with other symmetries 
can be mixed and we have to treat their spatial variations.
In order to do this, 
we diagonalize the mean-field Hamiltonian by solving the following
Bogoliubov de Gennes (BdG) equation\cite{dG}
\begin{equation} 
\sum_j 
\left [\begin{array}{clcr}
W_{ij}^{(\uparrow)} & F_{ij} \\
F_{ji}^* & -W_{ji}^{(\downarrow)}
\end{array}\right ]  u_{jn}= E_n u_{in} , 
\end{equation} 
where $E_n$ and $u_{in}$ are the energy eigenvalue and the 
corresponding eigenfunction, respectively.
The unitary transformation $\Psi_i = \sum_n u_{in}\Gamma_n$ 
diagonalizes the matrix ${\cal H}_{\rm MFA}$, and  
conversely the OPs ($\Delta_{ij}$, $\chi_{ij}^{(\sigma)}$  
and $\langle S^z_i \rangle$)  
can be written in terms of $E_n$ and $u_{in}$. 
These constitute the self-consistency equations which  
will be solved numerically in the following.\cite{Nishi}
We use $J$ as a unit of energy (i.e., $J = 1$), and take $t/J = 3$ 
throughout in this paper. 

First we examine the effect of spin-exchange scattering. 
The potential scattering is also taken into account,  
and the value $V_0 = 100$ is used. 
For this value of $V_0$ the scattering is in the unitary limit, 
since calculated hopping OPs ($\chi$) connected to the impurity site 
vanish, indicating that electrons cannot hop to the impurity site. 
The spatial variations of the OPs for $J_0 =1$ are shown in Fig.1.
It is seen that $\Delta_d$ is suppressed near the impurity (Fig.1(a)),  
and the effect is strongest along the diagonals of the square lattice. 
This is due to the interference effects for momenta close to the 
gap nodes. 
In the region where $\Delta_d$ is not uniform 
the extended $s$-wave OP ($\Delta_s$;  not shown) and the staggered 
magnetization (Fig1(b)) are induced. \cite{Salkola}
The important point here is that the spin-triplet SCOPs, 
$\Delta_{px}$ and $\Delta_{py}$,  can also appear (Fig1(c); 
$\Delta_{py}$ has a similar behavior as $\Delta_{px}$).
$\Delta_{px(y)}$ is defined as 
$\Delta_{px(y)}(i) = (\Delta^{(T)}_{i,i+x(y)} - \Delta^{(T)}_{i,i-x(y)})/2$ 
where $\Delta^{(T)}_{ij} = (\Delta_{ij} - \Delta_{ji})/2$ is 
the spin-triplet OP. 
In the presence of the megnetization there is 
the imbalance of the densities of spin-up and spin-down electrons. 
Then electron pairs cannot be formed in singlet channels 
only, and the spin-triplet components appear. 
We have also examined other values of $J_0$. The results are 
qualitatively the same, and the effect of spin-exchange scattering becomes
larger (smaller) with increasing (decreasing)  $J_0$ as expected. 

The above results can be understood more precisely using the 
Ginzburg-Landau (GL) theory. 
(The GL theory is not quantitatively valid except near $T_{\rm c}$, 
but it can give qualitatively correct results.) 
The GL free energy in the continuum representation 
can be written as\cite{SigUe}
\begin{equation} 
\begin{array}{rl}
{\cal F}_S   = & \displaystyle \int d^2r \Bigl(\sum_{j=d,s,px,py} 
\big[\alpha_j|\Delta_j|^2 + K_j|{\vec \partial}  \Delta_j|^2 \\
& \\
+ & \displaystyle g_j |\Delta_j|^2 \delta(r) \big] \\
& \\
+ & \displaystyle K_{ds} \big[(\partial_x \Delta_d)(\partial_x \Delta_s)^* 
- (\partial_y \Delta_d)(\partial_y \Delta_s)^* + c.c.\big] \\
& \\
+ & \displaystyle K_{dp} \big[\Delta_d \big\{(\partial_xm)\Delta_{px}^*-
(\partial_ym)\Delta_{py}^*\big\}m  + c.c\big] \\
& \\
+ & \displaystyle K_{sp} \big[\Delta_s \big\{(\partial_xm)\Delta_{px}^*+
(\partial_ym)\Delta_{py}^*\big\}m + c.c\big].  \\
%& \\
%+ & \displaystyle \big[\sum_{j=d,s,px,py} g_j |\Delta_j|^2 
%- g_m m^2\big] \delta(r) \Bigr). 
\end{array}
\end{equation}
This ${\cal F}_S$  is invariant under all symmetry  operations for the square 
lattice and we have dropped higher order terms.
Here we assume that all coefficients in ${\cal F}_S$ are positive 
except $\alpha_d$. 
Due to the $g_d$ term $\Delta_d$ is suppressed and its gradient becomes 
finite over the range of the coherence length.  Then $\Delta_s$ is induced 
through the mixed gradient ($K_{ds}$) term. 
The magnetization (denoted as $m$) is assumed to be induced by 
the spin-exchange interaction and have staggered oscillations. 
When $m$ and $\Delta_d$ coexist and have 
spatial variations, $p$-wave OPs can occur due to $K_{dp}$ term. 
In Fig.1 $\Delta_{px(y)}$ is finite where $m$ is finite, and this result 
is consistent with the above argument.
All induced OPs are real, since they are determined by bilinear coupling terms 
in ${\cal F}_S$.
Note the proximity effect in a  bilayer system composed of a $d$-wave 
superconductor and a (anti)ferromagnet can similarly induce 
$p$-wave OPs.\cite{prox}

Next we consider the spin-orbit interaction. 
It was argued, based on a continuum model that 
the spin-orbit impurity scattering in a $d_{x^2-y^2}$-wave superconductor 
can locally create a ($d_{x^2-y^2} + id_{xy}$)-state which breaks time-reversal 
symmetry(${\cal T}$).\cite{Balatsky,Graf,Simon}  
This is because the ($d_{x^2-y^2} + id_{xy}$)-state has an orbital moment 
and its coupling to the impurity spin can lower the energy of the system.
We will show that in the $t-J$ model a similar ${\cal T}$-breaking state
can also appear. 
In order to see this we add small $t'$ and $J'$ terms, since 
the $d_{xy}$-component ($\Delta_d'$)  is the SCOP defined on the  
next-nearest-neighbor bonds. 
(Note $J'$ of the order of $J$ is necessary to have finite $\Delta_d'$ 
in a system without impurities.) 
The spin-orbit coupling for Ni is estimeted to be 
$g/a \sim $30 meV ($a$ being the lattice constant).\cite{Yosida,Graf} 
Here we use a larger value $g/a  = 2.4 J$ in order to compensate the fact 
that the long-range part of $H_{so}$ is neglected in our calculation. 
($J$ is estimated to be $J \sim 0.13$ eV, so our parameter is 
about 10 times larger than the realistic value.) 
We find that $\Delta_d'$ (Fig2(a), (b)) as well as $\Delta_s$ is actually 
induced, but not $\langle S^z \rangle$ and $\Delta_p$.
The induced OPs have nontrivial phase structure, 
namely the phases take values other than 0 or $\pi$ 
(relative to $\Delta_d$) and change as functions of the position. 
In particular $\Delta_d'$ has a phase $\pi/2$ 
along the diagonal direction where $\Delta_d$ is strongly suppressed. 
This leads to a spontaneous current around 
the impurity (Fig.2(c)),\cite{Okuno} and the 
gap nodes in the $d_{x^2-y^2}$-wave SC state vanish due to the complex 
combination of SCOPs. 
These results are consistent with those in ref.19-21.\cite{ds}
For the parameters used here the magnitude of the current is of the order 
of 10 nA near the impurity and is smaller in other regions, and the gap due 
to the induced OP is $\sim 0.1$meV. 
A rough estimate using Biot-Savart law gives a value of $\sim$0.1 G for a 
magnetic field produced by the current at the impurity site. 

Now we examine the case where the spin-exchange and the spin-orbit 
interactions coexist. 
In this case we find a state which carries a spontaneous spin current (Fig.3) 
as well as usual charge current. 
The spin current is defined as the difference of the currents of spin-up 
and spin-down electrons. Its magnitude is typically $10^{-2}$ of 
that of the charge current. 
The occurrence of the spin current is naturally understood as follows. 
Spontaneous currents can flow because the SCOPs have nontrivial 
phase structure due to the spin-orbit coupling $g$. 
In the presence of the spin-exchange interaction $J_0$, the magnetization 
$m$ is finite, indicating that there is the imbalance of the densities of 
spin-up and spin-down electrons. 
Then the contributions to currents from electrons of opposite spins 
are not equal. Hence the spin current arises. 

In summary we have studied the effect of a magnetic impurity 
in the superconducting state of the $t-J$ model. The spin-exchange
scattering induces spin-triplet SCOP near the impurity, while the 
spin-orbit interaction leads to a state with complex OPs 
and a spontaneous current. When both interactions coexist 
we find a spontaneous spin current as well as the usual charge current. 
In the present work the long-range part of the spin-orbit interaction 
and the vector potential are not included, the latter of which has 
a feedback effect of the spontaneous current on the electronic 
states.\cite{neglect} 
Therefore the results presented here can be compared with 
experiments only in a qualitative sense. 
If some element with a larger spin-orbit coupling 
can be doped in high-$T_c$ superconductors, 
there would be a larger (compared with Ni) possibility to realize a SC 
state with a spontaneous current 
and a full gap in the local density of states. 
The former (latter) could be detected by $\mu$SR (low-temperature STM) 
measurements. 
				
The authors are grateful to M. Sigrist for useful discussions. 
We also thank T. Nishino and T. Hikihara for helpful advice on numerical 
calculations. 
This work was supported in part by a Grant-in-Aid for Scientific 
Research from the Ministry of Education, Science, Sports and Culture 
of Japan.

%\vfill\eject
%\centerline{\bf References} 

%\vfill\eject
%\centerline{\bf Figure Captions} 

\bigskip
\noindent 
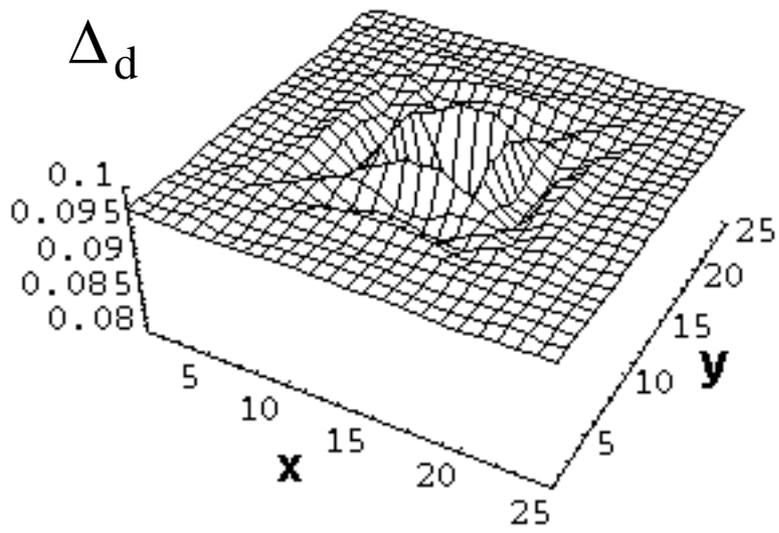
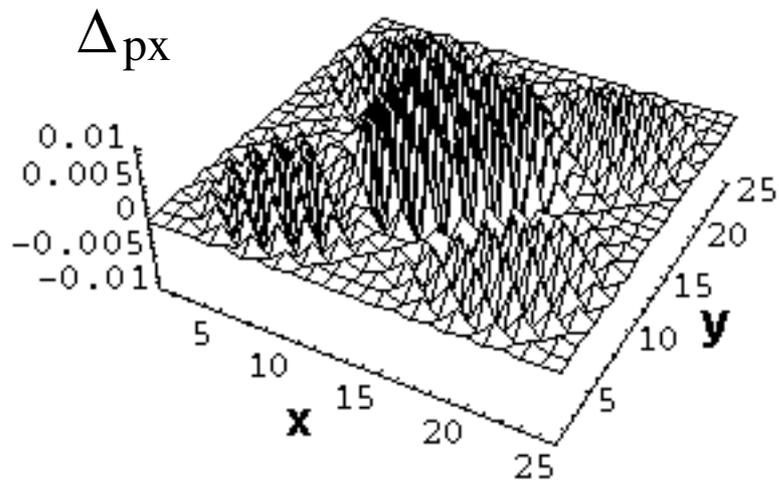
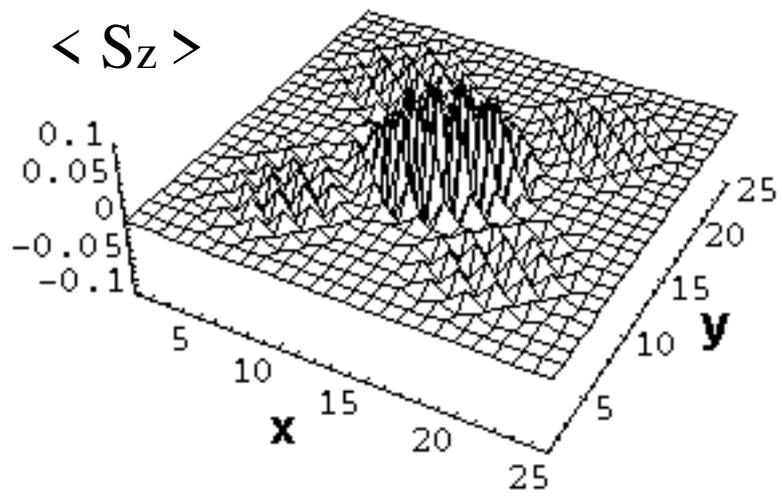
  {\bf Fig. 1}  
  Spatial variations of OPs for $t =3$, $J = 1$, $t' = J' = 0$, $\mu = -0.339$ 
  ($\delta=0.2$),  $V_0 = 100$, $J_0 =1$ and $g = 0$. 
  The system size is $25 \times 25$ sites, and an impurity is located at 
  the center of the system. 
  (a) $\Delta_d$,  (b) $\Delta_{px}$ and (c) $\langle S_z \rangle$. 
  Note that all OPs are non-dimensional.
  
\bigskip
\noindent 
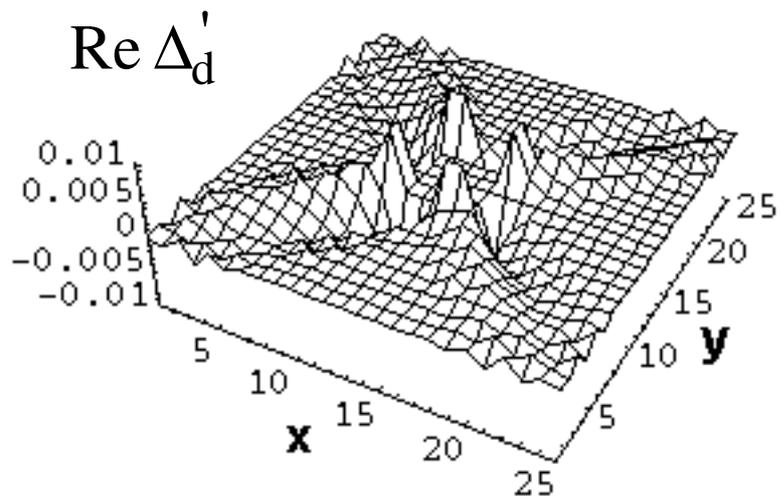
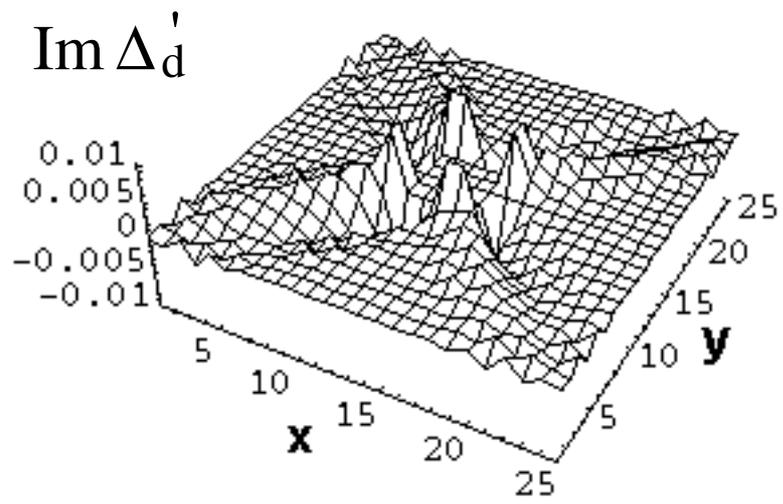
  {\bf Fig. 2} Spatial variations of (a) Re$\Delta_d'$, (b) Im $\Delta_d'$,  
  and (c) the spontaneous current around the impurity. 
  Parameters used are 
  $t =3$, $J = 1$, $t' = -0.5$, $J' = 0.2$, $\mu = 0.4$ ($\delta = 0.16$), 
  $V_0 = 100$, $J_0 =0$ and $g = 2.4$. The arrows in (c) indicate only the 
  directions of the currents, but not the magnitudes.

\bigskip
\noindent 
  {\bf Fig. 3}
  Spin current around the impurity for 
  $t =3$, $J = 1$, $t' = -0.5$, $J' = 0.2$, $\mu = - 0.5$ ($\delta = 0.189$), 
  $V_0 = 100$, $J_0 =0.6$ and $g = 4.8$. 
  The arrows indicate only the directions of the currents, 
  but not the magnitudes.


\begin{thebibliography}{99}
\bibitem{Gutz}
In this sense the slave-boson MFA at $T = 0$ is essentially 
equivalent to the Gutzwiller approximation, except the differences in 
renormalization factors.
\bibitem{Nishi} For numerical diagonalization of matrices we have used 
the computer program written by H. Nishimori; TITPACK version2.
\bibitem{ds}
Since OPs other than $\Delta_d$ and $\Delta_s$ are also mixed, 
the symmetry of the state is not exactly 
($d_{x^2-y^2} + id_{xy}$)-wave in contrast to ref. 19-21.
\bibitem{neglect} 
Besides the effects within the mean-field level, the magnetic field 
at the impurity site would weaken the Kondo effect, since the 
direction of ${\vec S}_{imp}$ would be fixed to the $z$-axis.
\end{thebibliography}
\end{document}